\documentclass[prl,twocolumn,superscriptaddress]{revtex4}
\usepackage{graphicx}
\usepackage{amsmath}
\usepackage{enumitem}
\newcommand{\be}{\begin{equation}}
\newcommand{\ee}{\end{equation}}
\newcommand{\ben}{\begin{eqnarray}}
\newcommand{\een}{\end{eqnarray}}

\newcommand{\affiliationSEAT}{Data Science Management and Market Intelligence, SEAT, S.A., Martorell, Spain}
\newcommand{\affiliationSTHAR}{Social Thermodynamics Applied Research (SThAR SA), EPFL Innovation Park B\^{a}t.~C, Lausanne, Switzerland}
\newcommand{\affiliationFQA}{Departament FQA, Facultat de F\'{\i}sica, Universitat de Barcelona, Barcelona, Spain}
\newcommand{\affiliationCCIT}{Centres Cient\'{\i}fics i Tecnol\`ogics (CCiT), Universitat de Barcelona, Barcelona, Spain}
\newcommand{\affiliationIFLP}{Instituto de Física La Plata-CCT-CONICET, Universidad Nacional de La Plata, La Plata, Argentina}
\newcommand{\affiliationIFISC}{Physics Department and IFISC-CSIC, University of the Balearic Islands, Palma de Mallorca, Spain}
\newlist{dlist}{enumerate}{1}
\setlist[dlist,1]{label=\arabic*.,leftmargin=0pt,itemindent=*}
\begin{document}

\title{Maximum Entropy Principle underlying the dynamics of automobile sales}

\author{D. Villuendas}
\affiliation{\affiliationSEAT}
\author{A. Hernando}
\email{ahernando@sthar.com}
\affiliation{\affiliationSTHAR}
\affiliation{\affiliationFQA}
\author{M. Sulc}
\affiliation{\affiliationSTHAR}
\author{R. Hernando}
\affiliation{\affiliationSTHAR}
\author{R. Seoane}
\affiliation{\affiliationCCIT}
\author{A. Plastino}
\affiliation{\affiliationIFLP}
\affiliation{\affiliationIFISC}

\begin{abstract}
We analyze an exhaustive data-set of new-cars monthly sales. The set refers to 10 years of Spanish sales of more than 
6500 different car model configurations and a total of 10M sold cars, from January 2007 to January 2017. We find that for those model configurations with a monthly 
market-share higher than 0.1\% the sales become scalable obeying Gibrat's law of proportional growth under logistic dynamics.
Remarkably, the distribution of total sales follows the predictions of the Maximum Entropy Principle for systems subject 
to proportional growth in dynamical equilibrium. We also encounter that the associated dynamics are non-Markovian, i.e., 
the system has a decaying memory or inertia of about 5 years. Thus, car sales are predictable within a certain time-period. 
We show that the main characteristics of the dynamics can be described via a construct based upon the Langevin equation. 
This construct encompasses the fundamental principles that any predictive model on car sales should obey.
\end{abstract}

\maketitle

The automobile industry is experiencing a deep economic and technological change. It is moving from (i) a fossil-fueled, private-ownership based,
manually-driven to (ii) an electric, sharing based, driver-less one\cite{pwc,mckinsey}. Understanding the supply-and-demand dynamics of automobile sales could help to achieve a smooth economic transition between (i) and (ii). Indeed, 88.1 million of both cars and light commercial vehicles were sold
worldwide in 2016\cite{macquarieresearch}, involving billions of dollars per year. Unplanned circumstances or forecast failures could  generate
immense loses in this market and its ties, as previously seen in past crisis and economic transitions as that after 2007\cite{wiki}.

Thanks to the rise of both  Big Data and digital tools we have today access to exhaustive databases that allow us
to analyze socioeconomic systems using ideas of statistical physics
\cite{sociophysics1,sociophysics2,sociophysics3}.  Examples run from allometric laws on city
populations\cite{cities1,cities2,cities3}, evolution of firm sizes\cite{firms}, transportation
networks\cite{transport1,transport3} and human mobility\cite{mobility}, to popularity of digital
products\cite{logistic}, the structure of the Internet\cite{net} or even diffusion of memes\cite{memes}.
In such an environment we analyze here the sales of new commercial vehicles in Spain from January 2007 to January 2017. The
corresponding exhaustive database is published by the Spa\-nish Directorate-General of Traffic (DGT) and contains registrations of more than 6\,500 different configurations (model + body shape + engine) and more than 10 million sold cars\cite{data}. To analyze the evolution of car sales, we apply, in particular, the procedure used in Refs.~\onlinecite{firms,entropy1,entropy2,entropy3} on the aggregated data of monthly registrations (or sales) per each car configuration (model) as provided by Ref.~\cite{data2}. The procedure is summarized as follows: 

\begin{dlist}
\item We first identify the main dynamical variables. In our case, the variable is the total number of  cars 
sold for the $i$-th automobile model at time $t$, $x_i(t)$.

\item We compare $x_i(t)$ versus its time derivative $\dot{x}_i(t)$ so as  to find indications of any
 possible  scaling rule in the underlying equation of motion,  of the form
\begin{equation}\label{eq:1}
\dot{x}_i(t) = v_i(t) [x_i(t)]^q,
\end{equation}
where $v_i(t)$ is the growth rate at time $t$ and $q$ is the exponent that parameterizes the dynamics. Due to the
stochastic nature of the growth rates, it is convenient to compare $x_i$ with the variance of $\dot{x}_i$, obtained from
all those $i$'s endowed  with a similar value of $x$. Thus, we fit the variance to an expression of the form
$\text{Var}[\dot{x}]= T_qx^{2q}$, where $T_q$ is used to measure the size of the fluctuations, that might be called  a
\textit{``temperature''}\cite{thermo2,thermo3}. If one finds contributions of several \textit{independent} components with the
form of Eq.~(\ref{eq:1}) with different exponents ---as $\dot{x}_i(t) = \sum_q v_{q,i}(t) [x_i(t)]^q$--- we will define
a temperature associated to each term from the fit $\text{Var}[\dot{x}]= \sum_q T_{q}x^{2q}$.

\item \label{ref:step3} Next, we independently analyze the density distribution of total  car sales per model $p(x)dx$. The principle of
Maximum Entropy (MaxEnt) with dynamical information states that the natural variable for measuring the entropy is the
one that linearizes the equation of motion, transforming the underlying symmetry into a translational
one\cite{MaxEntDyn}. For an equation of motion as Eq.~(\ref{eq:1}), this is achieved via the Tsallis
logarithm and its inverse function the Tsallis exponential\cite{tsallis} defining the new variable $u = \log_q(x/x_0)$ ---or $x=x_0\exp_q(u)$--- where $x_0$ is any value of reference (needed to keep the arguments dimensionless). After this transformation, the equation of motion becomes $\dot{u_i}(t) = v_i(t)$, which is
indeed translationally invariant. If the system is in equilibrium, the most probable distribution is the one that
maximizes the entropy under the system's observable constraints. If the form of the empirical distribution fits the form
predicted by MaxEnt, we can perform an independent quantitative measure of $q$ directly from the equilibrium
distribution.

\item If the value of $q$ one gets from the equation of motion, and the value of $q$ measured from the distribution-fit
agree, we assume that our procedure is correct, proving that the system is in dynamical equilibrium and obeys the
Maximum Entropy Principle.

\item \label{ref:step5} As an additional test for our procedure, we also perform
numerical experiments by defining microscopic equations of motion equivalent to
those observed in the empirical system. We then analyze the evolution of the
simulated system following the same steps as we traversed  for the empirical
one. If we obtain equivalent results, we would reconfirm that our procedure is
indeed correct.
\end{dlist}

\begin{figure}
\includegraphics[width=0.45\textwidth]{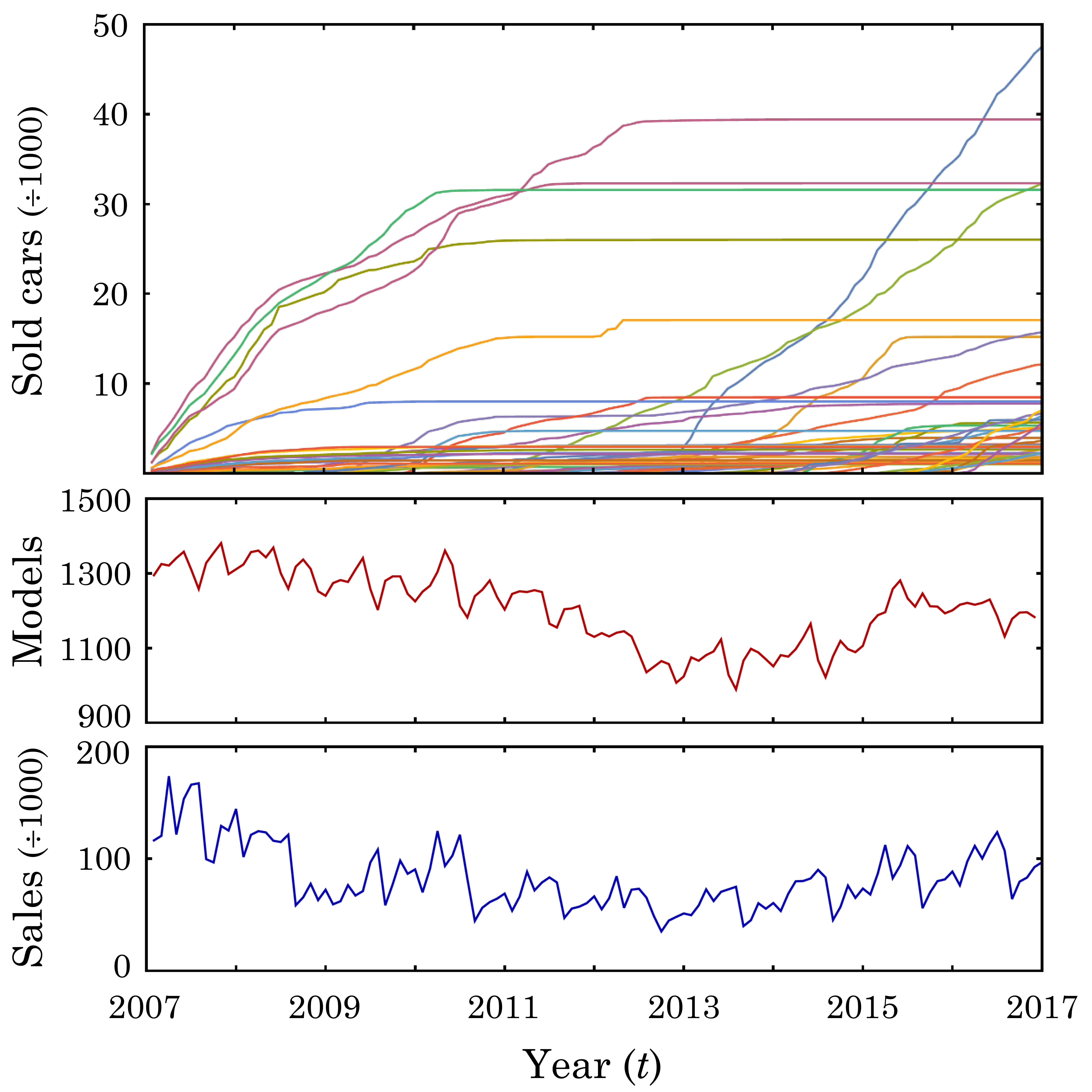}
\caption{\label{fig1}
    Top: Evolution of  car sales for a sample of car-models of the database. Saturation (as that described by the
    logistic equation) is observed. Middle: Evolution of the number of different models available per month during the
    ten years recorded in the database. Bottom: Monthly new car sales in Spain, with a total of $10\,062\,193$ sales
    during the last ten years.
}
\end{figure}

As stated in step 1, our relevant variable is the total of  car sales $x_i$ for the $i$th automobile model. We display
in Fig.~\ref{fig1} a sample of some trajectories for the ten years covered by our dataset, where a logistic profile is clearly
visible (i.e., growth with saturation). Saturation occurs when the popularity of a model drops because of the
availability of a newer version, or if a competitor advances on  the same market niche. We also show in Fig.~\ref{fig1}
the evolution of the number of different models sold per month during the ten years included in the dataset.

\begin{figure}
\includegraphics[width=0.45\textwidth]{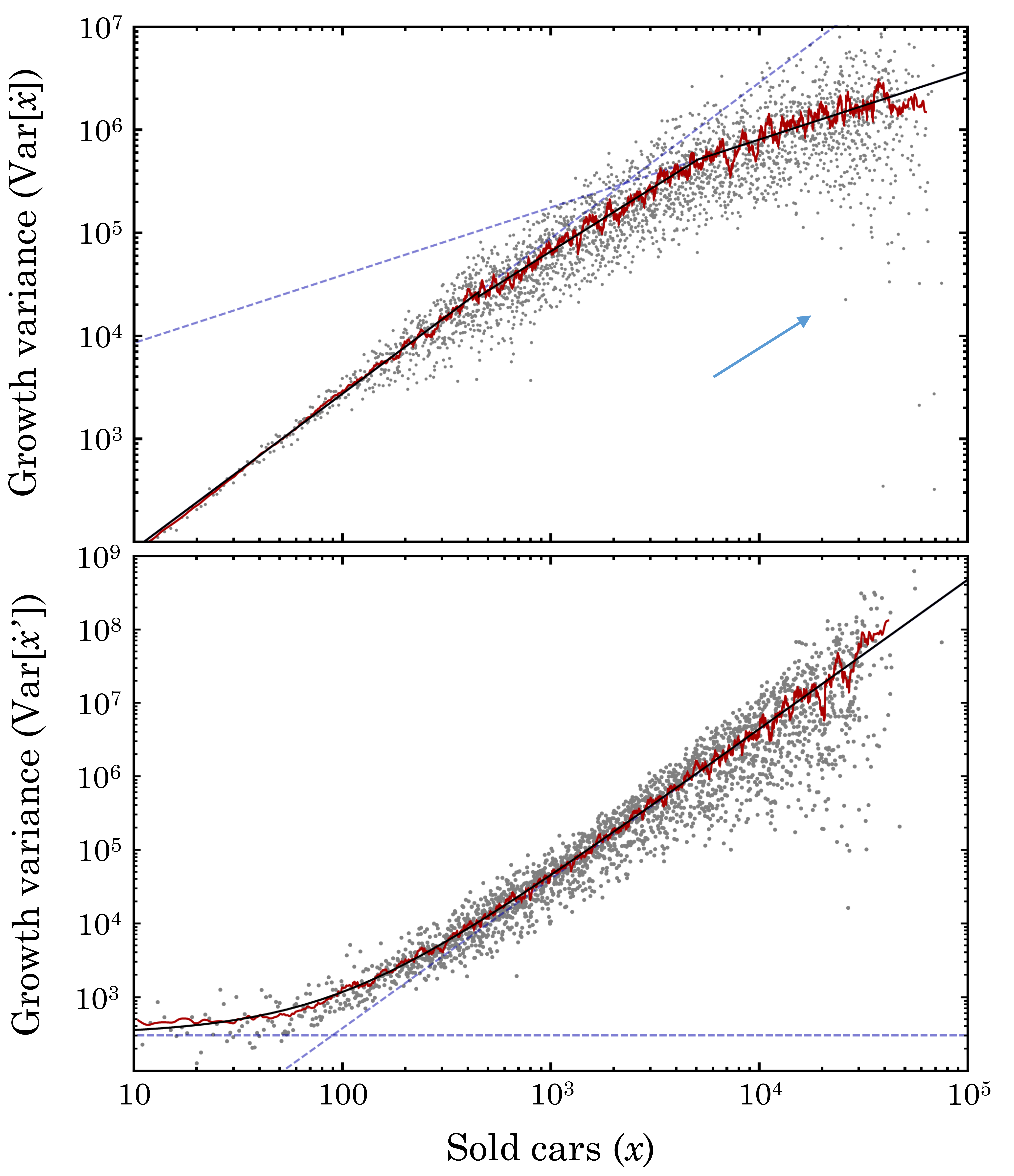}
\caption{\label{fig2}
Top: Empirical variance of the sales' growth versus total  car sales (gray points). We compare the moving average
(red line) with the analytical fit to  Eq.~(\ref{eq:1}) (black line). We find three regimes with different
exponents. Bottom: same as top, but with growth corrected by the logistic equation (see text), that fits to the form
of Eq.~(\ref{eq:2}).
}
\end{figure}

When we compare $x$ vs. $\text{Var}[\dot{x}]$ so as to fit Eq.~(\ref{eq:1}), we find what looks like a three-component
equation of motion, with an exponent i) for low sales of $q=0.75\pm0.01$, ii) for mid-sales of $q=0.63\pm0.01$, and iii)
for high sales of $q=0.33\pm0.01$. R2 adopts a  value of $0.9997$ (see Fig.~\ref{fig2}). A naive interpretation might be
that Eq.~(\ref{eq:1}) is indeed the underlying equation of motion. The problem would be to explain the origin of the
values for three exponents mentioned above. Even worse, an independent assessment of $q$ derived  from the equilibrium
distribution will not match these 3 values, as we will show later. However, paying attention to the high sales in
Fig.~\ref{fig2} (arrow), we appreciate the existence of points with low variance. Indeed, here growth rates drop to such
an extent that one approaches saturation and this would generate an underestimation of the exponents. To get deeper insights into the
actual underlying dynamics, we need to correct for each car
model the effects of saturation. To this aim, we need to reconsider the form of the logistic equation\cite{logistic}
\begin{equation}
\dot{x_i}(t)=v_i(t) x_i(t)\left[1-x_i(t)/X_i\right],
\end{equation}
where $X_i$ is the final  number of total car sales. If we reconsider the trajectories by first defining
$\dot{x_i}'(t)=\dot{x_i}(t)/[1-x_i(t)/X_i]$, we should recover an expression of the form of Eq.~(\ref{eq:1}) as $\dot{x_i}'=v_i
x_i$ that corrects growth rates for the effects of saturation. Comparing in this way $x$~vs.~$\text{Var}[\dot{x}']$, we obtain the
curve displayed in Fig.~\ref{fig2}, which nicely describes a three-terms function: one has $\text{Var}[\dot{x}'] = T_1
x^{2q} + T_{1/2} x + T_0$, with $q=1.016\pm0.002$, and an R2 coefficient equal to $0.9998$ [$\log(T_1)=-3.41\pm0.05$,
$\log(T_{1/2})=1.60\pm0.06$,  and $\log(T_0)=5.73\pm0.06$]. This new result vouches for an equation of motion of the
form
\begin{equation}\label{eq:2}
\dot{x}_i'(t) = v_{1,i}(t) x_i(t) + v_{1/2,i}(t) \sqrt{x_i(t)} + v_{0,i}(t).
\end{equation}
These three components have been already observed for firm sizes\cite{firms}, city populations\cite{entropy2}
and are well-understood. The first term with $q=1$ is Gibrat's law of proportional growth\cite{gibrat},
which is expected for multiplicative processes. Such behavior indicates that the popularity of an automobile model grows
as more cars are sold, following a rich-get-richer mechanism. The second term, with exponent $q=1/2$, emerges from the
proportional growth as a side effect of the central limit theorem (or finite-size effects), as shown in
Ref.~\onlinecite{entropy3}. It has been shown that this component is characterized by uncorrelated noise and no extra physics
is expected. Finally, the last term in Eq.~(\ref{eq:2}) correspond to linear forces ($q=0$), independent of the value of
$x$. This last term becomes relevant only for small number of sales.

\begin{figure}
\includegraphics[width=0.45\textwidth]{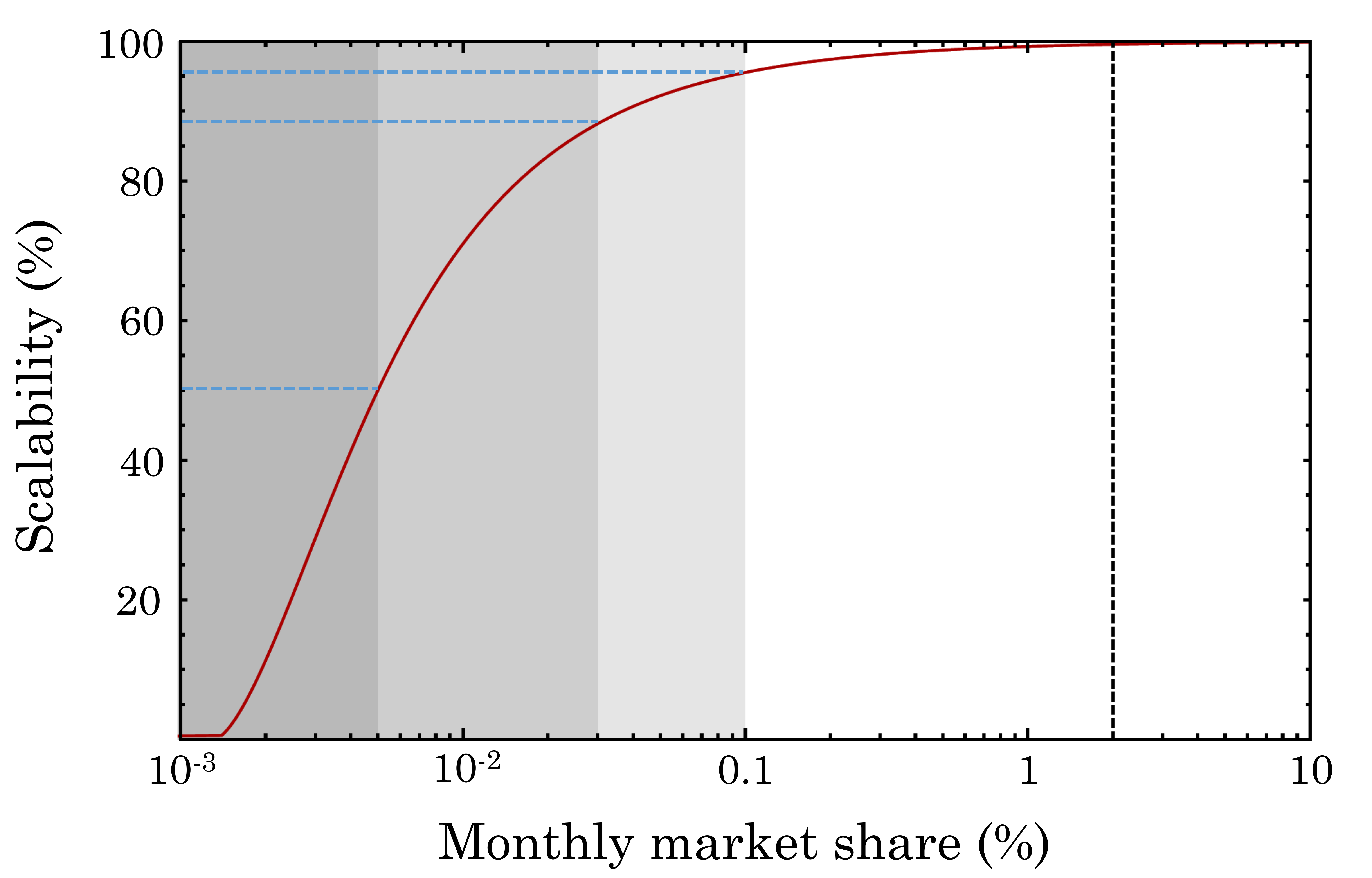}
\caption{\label{fig3}
Scalability (defined as the contribution of the proportional term to the total growth in Eq.~(\ref{eq:2})) as a
function of the monthly market share: with 0.005\% of the market share, the proportional term starts to be
dominant, reaching around 90\% of the total contribution with 0.03\% of  monthly sales and a 95\% with 0.1\% of
the market (dashed blue lines and shaded areas). The average maximum market share for a single model is around
2\% (dashed gray line).
}
\end{figure}

In our procedure we will focus only on the proportional regime, where the most  interesting physics takes place. However,
one may ask:  which is the critical number of sold cars so as to consider that the pertinent trajectory would be
accommodated by the proportional regime? A first estimation can be obtained via the overlap between the first and second
terms of Eq.~(\ref{eq:2}), obtaining $x_0 = T_{1/2}/T_1 = 129$ cars. At this critical value we have equal contributions
from each of the two terms. It is then natural to ask when we can consider that the dynamics is {\it fully} governed by
proportional or scalable growth. We can combine our two previous fits to compare the market share or relative number of
sold cars per month $\chi=\dot{x}/\sum_{j}{\dot{x_j}}$ with the relative contribution of the proportional term to the
total growth, that we call {\it scalability}: $\kappa(x) = \sqrt{T_1 x^2/\text{Var}[\dot{x_i}']}$. As seen in
Fig.~\ref{fig3}, an automobile model is $\kappa=95\%$ into the scalable growth regime  when it reaches a $\chi=0.1\%$
of monthly relative sales or market-share.

\begin{figure}
\includegraphics[width=0.45\textwidth]{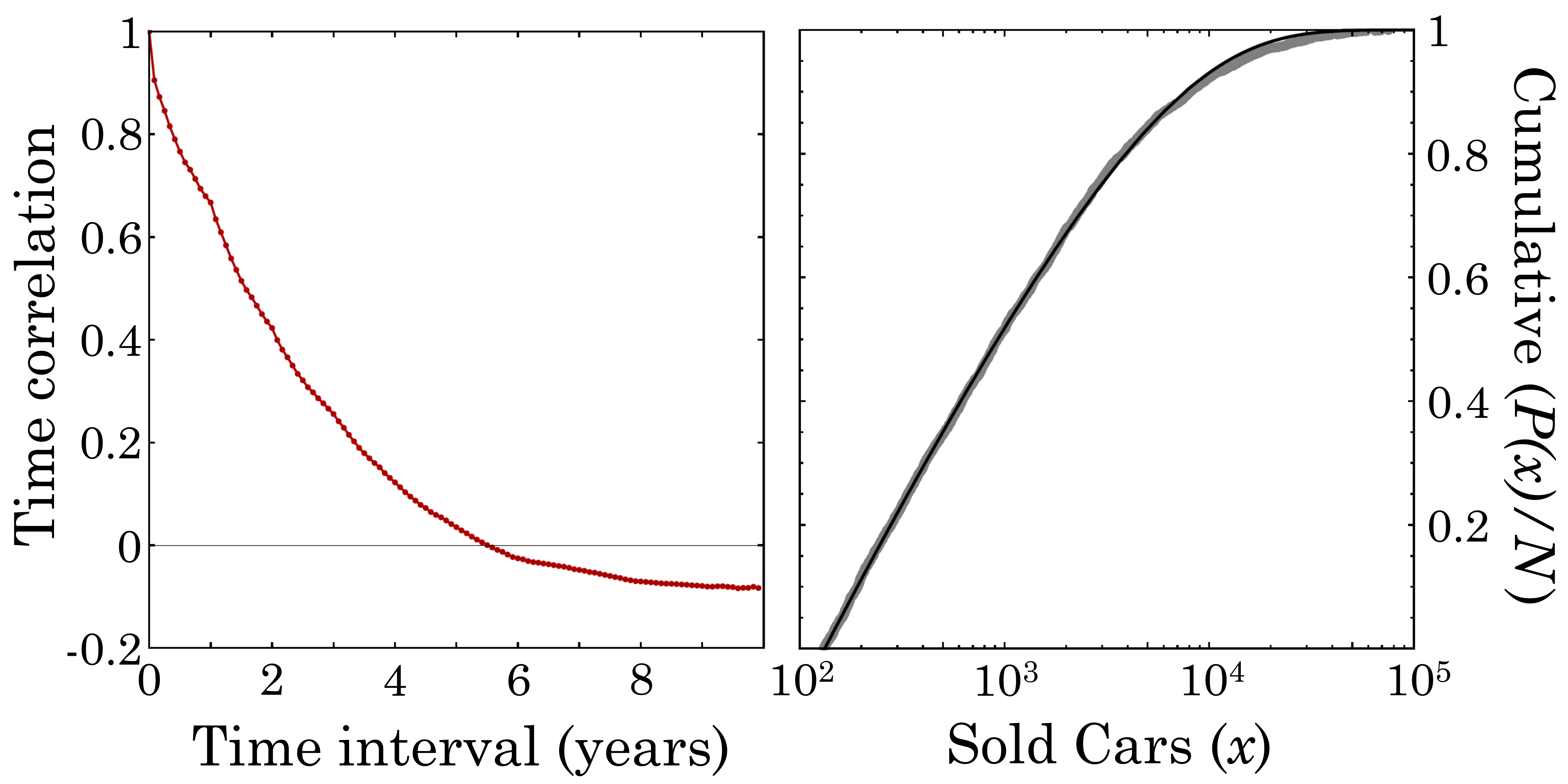}
\caption{\label{fig4}
Left: Time correlation of the growth rates as a function of the time interval. A clear decay is found that becomes
slightly negative after 5 years. Thus, predictability is possible, but for  no longer than  5 years. Right:
Empirical cumulative distribution of total sold cars per model (gray dots) compared with the MaxEnt prediction
[see Eq.~(\ref{eq:7})] (black line).
}
\end{figure}

Another relevant question regarding growth rates becomes legitimate at this point. It is crucial for predictability: Are
car sales a non-Markovian process? In other words,  are the sales of today correlated with those of yesterday, and thus
predictable? Buying a car is an individual decision, undertaken according to a plethora of various considerations. We
are looking at a complex decision-process that takes its time and relies on decisions taken previously by other
individuals. As we did for cities' populations in Refs.~\onlinecite{cities3,memory} which unraveled the existence of a
``memory'' in cities, we have attempted to define here the putative system's memory as the time correlation
$c(\tau)=\text{Cor}[\dot{x}(t),\dot{x}(t+\tau)]$, where $\tau$ is the correlation time interval. We find an exponential-like decay, with slightly negative values after
$\tau= 5$ years as shown in Fig.~\ref{fig4}. This in turn establishes the limits of predictability: no accurate prediction on sales can be done for
longer than that time-period.


We continue now with the step~3 outlined on page~\pageref{ref:step3}. For  maximizing the entropy $S$, it was shown in
Ref.~\onlinecite{logistic} that the only observable and objective constraints for systems with logistic growth are, due to
the limitation in units of resources, the total available units (in our case, total number of sold cars $X$) and the
number of elements sharing these units (for us, the number of available automobile models $N$). Considering only those
car-models the sales of which exceeded $x_0=130$ units in the ten years (the limit of proportional growth as derived before) we
have $X=9\,957\,537$ and $N=3\,084$. Following Refs.~\onlinecite{firms, entropy1,entropy2,entropy3}, we write our
thermodynamic potential as:
\begin{equation}\label{eq:3}
\Omega = -T\,S - \mu N + \Lambda X
\end{equation}
where $T$, $\mu$, and $\Lambda$ are the concomitant Lagrange multipliers for the general variational problem of the
thermodynamic potential $\Omega(S,N,X)$. For a general equation of motion with arbitrary $q$, we assume that an
underlying probability density $p(u)$ governs our process, where $u=\log_q(x/x_0)$, $N = \int_0^\infty du p(u)$ and $X = \int_0^\infty du p(u) \exp_q(u)$. We explicitly cast the MaxEnt problem as a $p-$variational one:
\begin{equation}\label{eq:4}
0 = \delta_p\Omega = \delta_p\!\!\int_{0}^\infty\!\!\!du p(u) \left\{ T \log[p(u)/N] -\mu -\Lambda \exp_q(u) \right\}
\end{equation}
where $\delta_p$ represent variations with respect to $p(u)$. We find the solution $p(u) = zN\exp[-\Lambda^*\exp_q(u)]$ which in terms of the variable $x$ is a power law with exponential cut-off:
\begin{equation}\label{eq:5}
p(x) = zN\frac{\exp(-\Lambda^* x/x_0)}{x^q},
\end{equation}
where $\Lambda^*=\Lambda/T$ and $z=\exp(\mu/T)$. The first constraint normalizes $p(u)$ to the number of car models $N$, and the second to the total  car sales, yielding the equations of state:
\begin{subequations}\label{eq:6}
\begin{align}
z &= \left( x_0\Lambda^* \right)^{1-q}/\Gamma(1-q,\Lambda^*),\\
x_0 N/X&=\Lambda^*\Gamma(1-q,\Lambda^*)/\Gamma(2-q,\Lambda^*).
\end{align}
\end{subequations}
The cumulative distribution can be written as
\begin{equation}\label{eq:7}
P(x)/N = 1-\frac{\Gamma(1-q,\Lambda^*x/x_0)}{\Gamma(1-q,\Lambda^*)}.
\end{equation}
Passing now to step 4 of page 1, we show in Fig.~\ref{fig4} the comparison of the MaxEnt prediction with the empirical
cumulative distribution of total  car sales. We have considered only sizes larger than $x_0=130$ and  fit
Eq.~(\ref{eq:7}) via $\log(x_0)$, $q$, and $\log(\Lambda^*)$ (logarithms are used for numerical stability).
We find a remarkable fit with $\log(x_0)=4.884\pm0.001$, $q=1.048\pm0.001$, and
$\log(\Lambda^*)=-4.560\pm0.003$. Our R2 value is $0.99993$ ---the relative
difference with the empirical values $\log(130)$, $1$, and $-4.8712$ (taken
from the values of $q$, $X$ and $N$) are $0.3\%$, $5\%$, and $8.8\%$,
respectively.
Thus, we can safely claim that {\it the dynamics of car
sales are very close to equilibrium and the distribution of total sales obeys the Maximum Entropy Principle for
scale-free systems}.

We finally proceed to step~5 on page~\pageref{ref:step5} by defining the equations needed  for performing microscopic simulations of  the
empiric system. For writing these equations, we need the following considerations: 

\begin{dlist}
\item Geometric Brownian walkers are known to obey Gibrat's law and reproduce states of dynamical equilibrium following
entropic laws. However, Brownian walkers are Markovian. We will use instead a Langeving-like equation with a viscous
term\cite{thermo2,thermo3}, which is known to reproduce time correlations with exponential decay.

\item The constraints in the thermodynamic potential of Eq.~(\ref{eq:3}) are introduced as forces in the Langevin
equation, in analogy to what is done in molecular simulations.

\item In addition to the proportional growth $q=1$, the model should include linear forces ($q=0$) and finite size
fluctuations ($q=1/2$). We include them as independent thermal baths with no inertia.

\item The number of sold cars can grow, but does not decrease (un-selling cars is not allowed). Thus, a Maxwell
demon\cite{thermo3} is introduced to select only positive values from the fluctuations;
\end{dlist}

\begin{figure}
\includegraphics[width=0.45\textwidth]{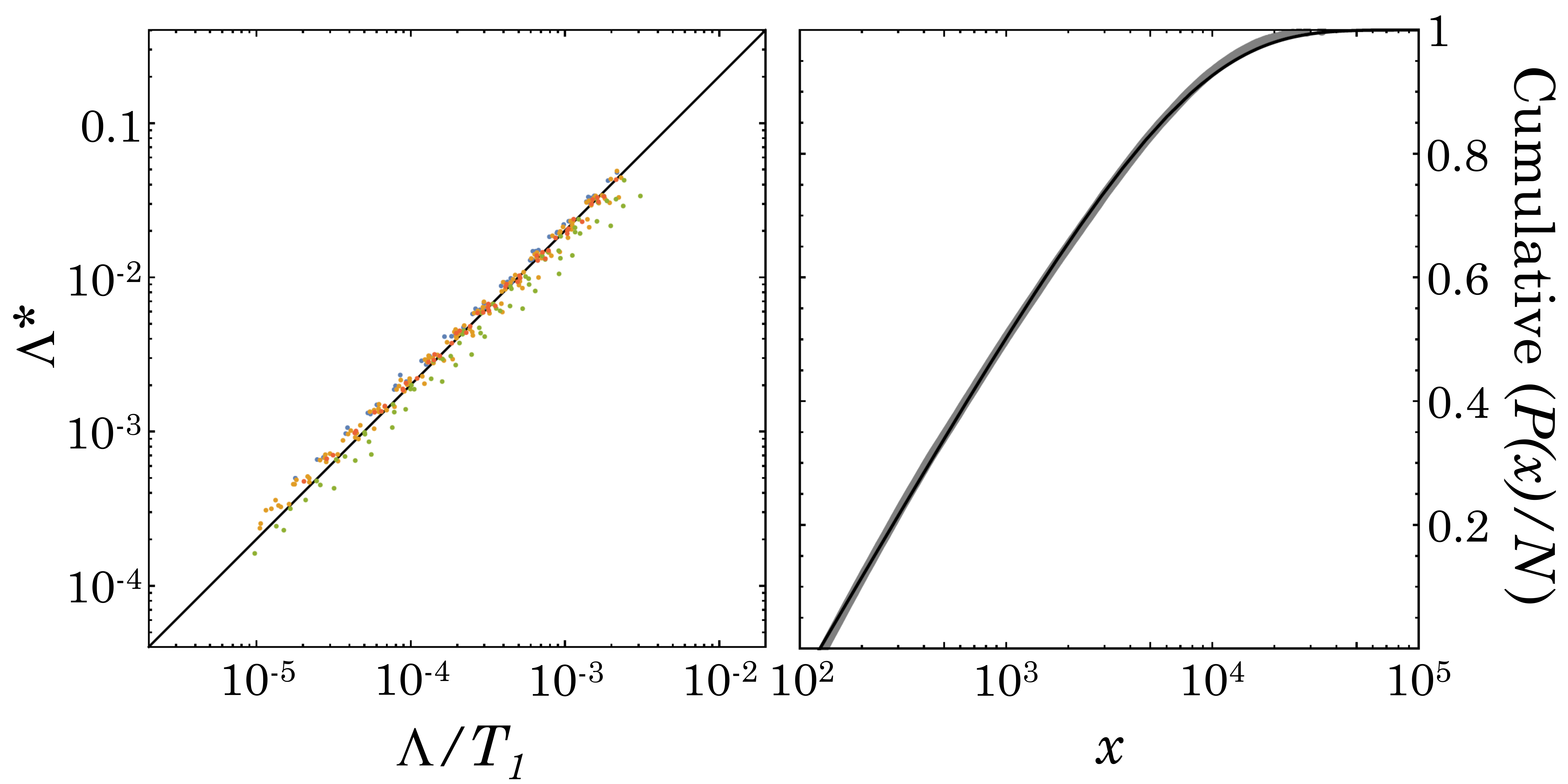}
\caption{\label{fig5}
Left: $\Lambda^*$ as measured from the equilibrium distribution versus the value of $\Lambda/T_1$ used in the
simulations. Each dot is an independent realization using the range of values for $\Lambda$ and $T_\eta$ described in
the text. Colors represent realizations with the same value of $\Delta t$.
}
\end{figure}

After these considerations, we write the full set of microscopic equations as:
\begin{align}\label{eq:8}
\dot{v}_{1,i}(t) &= \eta_i(t) - \Lambda x_i(t) - \gamma v_{1,i}(t), \\
\dot{x}_i(t) &= v_{1,i}(t) x_i(t) + v_{1/2,i}(t)\sqrt{x_i(t)} + v_{0,i}(t) \vee 0,\notag
\end{align}
where $\gamma$ is the dumping or viscosity which controls the inertia,   
and the thermal bath obeys $\langle\eta_i(t)\eta_j(t')\rangle = 2T_\eta\gamma~\delta_{ij}\delta(t-t')$ with $T_\eta$
defining its temperature. The other two sources of noise are described as White Noise with $\langle
v_{1/2,i}(t)v_{1/2,j}(t')\rangle = 2T_{1/2}/\Delta t~\delta_{ij}\delta(t-t')$, and $\langle v_{0,i}(t)v_{0,j}(t')\rangle
= 2T_0/\Delta t~\delta_{ij}\delta(t-t')$ with $T_{1/2}$, and $T_0$ their respective temperatures. Here, $\delta$ is the
Dirac's Delta and $\Delta t$ is the discretized interval of time used to numerically solve our equations. Finally, the
symbol $\vee$ indicates choosing  the maximum between the quantity on the left and on the right, preventing
negative values of the growth $\dot{x_i}$.

Since our aim is to prove that the entropic procedure properly predicts the equilibrium distribution in
the proportional regime, we have reduced Eqs.~(\ref{eq:8}) for the sake of simplicity by removing the finite size
term $q=1/2$ (as done in Ref.~\onlinecite{firms}) and considering over-dumping dynamics by setting $\gamma=1/\Delta t$.
(For the interested reader, a full exhaustive exercise solving and exploring Eqs.~(\ref{eq:8}) will be published
elsewhere.) 

We have solved the simplified version of Eqs.~(\ref{eq:8}) for the range of values $0.1\le T_\eta\le500$ and $10^{-4}\le
\Lambda\le0.5$ with $N=3000$, $T_0=100$, and intervals of times in a range of $10^{-3}\le \Delta t \le 10^{-2}$,
starting at an initial configuration where every walker has $x(0)=1$. Eventually, the growth rate of the walker becomes
zero due to the force induced by the $\Lambda$ term, reaching saturation. So as to  keep the number of active walkers
constant, we add a new one at $x=1$ every time an older walker saturates. After some time-intervals have elapsed, the
system reaches a thermodynamical equilibrium. 

We study the scalable growth by selecting only walkers with values $x_i(t)>100$ and measuring the effective temperature
as the variance of the growth rates $T_1 = \text{Var}[\dot{x}/x]$ at equilibrium. We find that the system's temperature
and that of the thermal bath obey $\log(T_1) = (0.52\pm0.06) + (0.65\pm0.02) \log(T_\eta)$. The equilibrium distribution
$p(x)dx$ fits the form of the MaxEnt prediction in Eq.~(\ref{eq:7}): when fitting via $q$ and $\Lambda^*$, we find
$q=1.01\pm0.01$ and $\log(\Lambda^*)=(3.0\pm0.2)+\log(\Lambda/T_1$) ---independently of the other parameters $\Delta t$
and $T_0$, see Fig.~\ref{fig5}--- confirming the correctness of our entropic procedure. We show in Fig.~\ref{fig5} the
cumulative distribution for a simulation using the empirical temperature $T_1=\exp(-3.41)$ ($T_\eta=0.0024$) and
$\Lambda=1.35\times10^{-5}$, obtaining $\Lambda^*=0.01$.

Summing up, we have shown that the dynamics underlying  car sales corresponds to a non-Markovian logistic growth with scale
symmetry, and the dynamical equilibrium obeys the maximum entropy principle. We have also found that the sales become
scalable up to a 95\% when a model reaches a 0.1\% of the monthly market share. Numerical experiments reproduce indeed
the macroscopic features of the empirical system and confirm the theoretical procedure proposed here.  In view of such
results, our macroscopic theoretical framework should be recommendable  in any simulation of car sales that aims to be
predictive. Today, the industry uses algorithms based on, for example, deep learning on time series that are very useful
to fit internal correlations between every involved variable. Our approach can be used to constrain the degrees of
freedom in such methods, which should further improve their efficiency. We hope that accurate forecasts will help to describe
controlled evolution of the market and to  reduce the economic risk inherent to profound technological transitions, such
as the one that the automobile market is nowadays experiencing.

\end{document}